# Thermal disorder in finite-length carbon nanowire

[1]C.H.Wong, [1]E.A.Buntov, [1]A.F.Zatsepin, [2]M.B.Guseva

[1]Institute of Physics and Technology, Ural Federal University, Yekaterinburg, Russia

[2]Faculty of Physics, Moscow State University, Moscow, Russia

**Abstract:**

Chemisorption is one of the active research areas in carbon materials. The occurrence of the monoatomic carbon chain can be made by surrounding the double walled carbon nanotube and meanwhile worldwide efforts have been made to create the extraction technique for unlashing the carbon chains from the enclosure. Here we report an extensive study of the kink structure in the free standing carbon nanowires. Our Monte Carlo simulation considers the multi-monoatomic carbon chains laterally interacted by the Van der Waal's force. Despite the linearity of the carbon nanowires is independent of chain length at low temperatures, the same situation does not hold at high temperatures. Disordered kink structure is observed in the short carbon chains especially above Peierls transition temperature. For instance, the average kink angle of 50-atoms carbon nanowire is as large as 35 degree at 800K. We have provided an important inspection that any physical property of the finite-length carbon chain predicted by ab-initio calculation should reconsider the atomic rearrangement due to the thermal instability. Apart from this, the kink structure in the nanowires likely increases the probability of attaching negatively charged atoms which is an encouragement to find the next generation materials for chemisorption.

**Introduction:**

Carbon exists in various configurations such as graphite, diamond, graphene, nanotube…etc. Graphene and its composites have been confirmed as promising adsorbents at high temperature environment such as hydrogen, carbon dioxide, methane in comparison to those traditional large surface area activated carbons [1]. The hydrogenated graphene contain sp$^3$ C-H bond on the basal plane made by chemical vapor deposition (CVD can be used as hydrogen storage [1]. Apart from this, dramatic enhancement of the absorbing selective organic molecule is also observed in the reduced graphite oxide foams [1]. Further modifying the structure into porous graphene oxide-like foam provides selectively absorption of carbon dioxide over nitrogen, methane, hydrogen and carbon monoxide at the conditions of 1x $10^6$ Pascal and 300K with help of the aliphatic and aromatic domains with oxygen rich functional groups on the surfaces [1]. The molecular absorption in the porous carbon nanotubes is also expected after the nanotubes are electrostatically charged in ionic solutions [2]. On the other hand, the curvature assisted effective atomic number Z is also proposed in the thinner nanotube which offers an alternative way to influence the electrostatic charges in the system [3]. If the carbon materials exist in the presence of numerous local curvatures, it may bring hopes for effective chemisorption especially for the negatively charged atoms due to the modified Z [2,4]. The new form of carbon, monoatomic carbon chain, becomes a hot research topic in recent years due to the theoretical giant elastic modulus, torsional induced magnetism and large electronic density of states at Fermi level [5,6,7]. The linear chained carbon contains two phases [8]. The metallic cumulene phase is likely more energetically favorable at low temperatures and the semiconducting polyvne phase should appear above the Peierls transition temperature at 500K [8,9]. However, the synthesis of the isolated infinite linear chain of carbon presents a huge challenge. Using the most modern technology could only obtain about 6000 atoms carbon nanowire which required protection from the double walled carbon nanotube unavoidably [10]. As the Boltzmann excitation is dominated at high temperatures [11], we are trying to investigate whether the kinks will be generated in the finite-length carbon nanowires upon heating. The disordered carbon chain alongside the kink structure may open an opportunity for design the next generation materials for chemisorption.

**Computational method:**

Based on the Monte Carlo simulation of 10 parallel monoatomic carbon chains in form of hexagonal array, the distribution of the bond angles will be computed in a series of chain lengths in the present of weak Van der Waal's force via the Hamiltonian $H$ below.

$$H = e^{-T/T_{bj}} \sum_{m=1}^{M} \sum_{n=1}^{N-1} \left| E_{m,n,j} e^{-\frac{(r_{m,n} - r_{m,n,j}^{eq})}{0.5 r_{m,n,j}^{eq}}} - E_2 \right| + e^{-T/T_{bj}} \sum_{m=1}^{M} \sum_{n=1}^{N-1} J_A (\cos\theta + 1)^2 - 4\varphi \left[ \sum_{n,m} (\frac{\sigma}{r})^6 - (\frac{\sigma}{r})^{12} \right]$$

where $M$, $N$, $E_2$, $T$ are the total number of chains, the total number of carbon in each chain, double bond energy and temperature respectively. The formation in single, double and triple bond corresponds to $j = 1, 2$ and $3$ respectively. The $j$ is a stochastic variable in the simulation. The $r$ is computed in Cartesian coordinate and $r_{m,n,j}^{eq}$ is equilibrium position. The temperature to break the covalent bond $T_{bj}$ is determined by $E_j = k_B T_{bj}$ where Boltzmann constant $k_B = 1.38 \times 10^{-23} JK^{-1}$. The Van der Waal's energy $E_{vdw}$ is the only interaction between the adjacent carbon chains with the sample length $\tau_s$. The $\varphi$ and $\sigma$ are $8.1 \times 10^{-23} J$ and $1.23 \times 10^{-10} m$ respectively [11]. The algorithm and other simulation parameters can be found in C.H.Wong etal [12]. The Van Der Waal's distance d is 0.3nm unless otherwise specified. Three adjacent carbon atoms form one pivot angle with the appearance of the kink structure eventually where the pivot angle in the linear carbon chain is defined as zero degree. The carbon chain carrying M atoms is named as M-CNA. For example, the carbon nanowire made of 50 atoms is named as 50-CNA.

**Results and discussion:**

Figure 1 shows the distribution of the kink angle of the carbon nanowires array at 1000K. Each chain consists of 2000 atoms and the lateral chain-to-chain distance is 0.3nm. The most probable kink angle is 1 degree. The appearance of the large kink angles is also observed but the probability leads to the exponential-like drop. The alignment of the 2000-CNA can be considered as linear because the average kink angle of the entire sample is only 2.9 degree. However, the situation is completely different in the short carbon nanowire. Figure 2 illustrates the average kink angle of the carbon nanowire in different chain lengths. The rise of the kink angle becomes an obvious feature if the nanowire contains less than 250 atoms. The average kink angle of 38 degree is attained in the 50-CNA as shown in Figure 2. The reason is that the short carbon nanowires are suffered from the lack of atomic spring constants and hence controlling the kinematic of any fast moving atom is not easy at high temperatures. In contrast, any atom displaced from its equilibrium position along the infinitely long carbon chain has to against numerous atomic spring constants in parallel simultaneously [11].

The ten disordered 50-CNA at 1000K is plotted in figure 3. In comparison to 2000-CNA, the broad variety of kink angles is noticed in the 50-CNA. The appearance of the kink angles bigger than 100 degree is rare because the formation of the large kink angle is impeded in order to minimize the electrostatic repulsion owing to the kink [4]. The inset of Figure 3 displays the atomic distributions at equilibrium correspondingly. Thermal energy actuates the lateral vibration of atoms phenomenally so that aligning the carbon atoms linearly at 1000K is prohibited. Another feature of the inset is that the lateral fluctuations are more obvious towards left hand side. It is originated from the one-end fixed boundary condition.

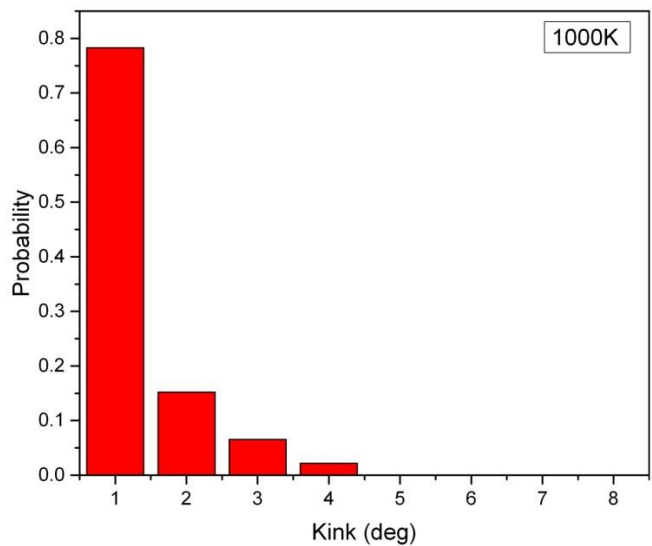

Figure 1: The distribution of kink angles in the 10 parallel 2000-CNA at 1000K. Each nanowire carries 2000 atoms

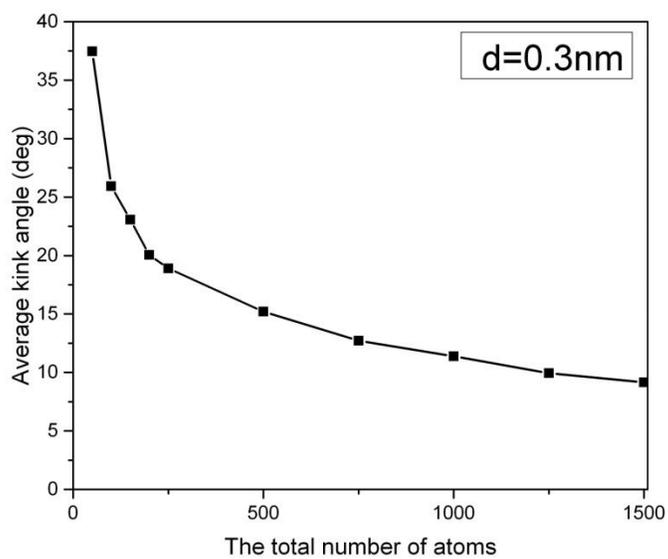

Figure 2: The average kink angle of the carbon nanowires array as a function of chain lengths at 1000K.

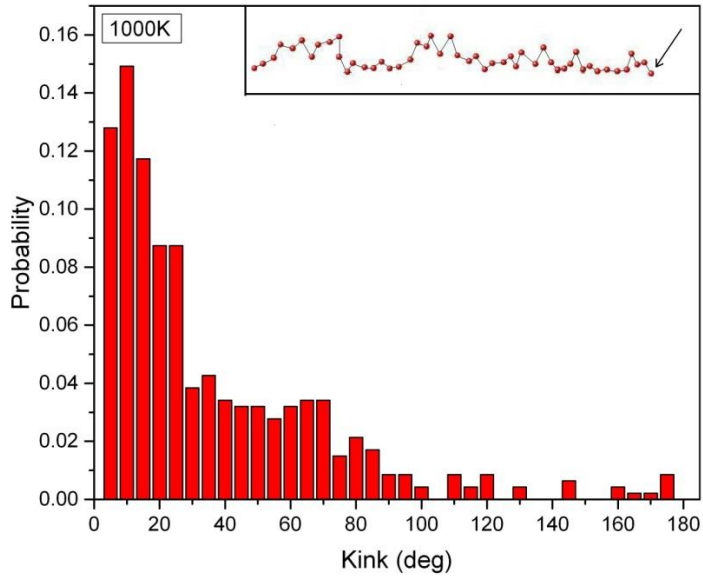

Figure 3: The distribution of kink angles in the 10 parallel 50-CNA at 1000K. Each nanowire contains 50 atoms. The one-end fixed boundary condition in the inset is marked by arrow.

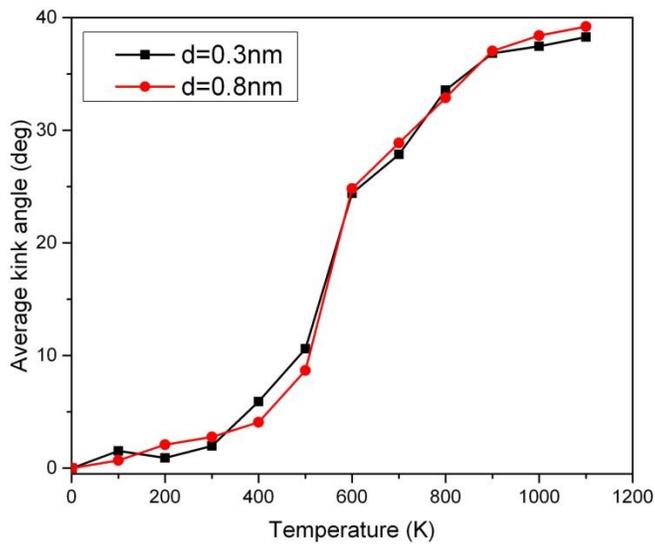

Figure 4: The thermal effect of the average kink angle of the 10 parallel 50-CNA in various Van Der Waal's couplings.

Using temperature sweep analysis, we have created a new method to probe the Peierls transition [8] of the short carbon chains by considering the slope or inflection point in Figure 4. No significant disorder in the 50-CNA below the Peierls transition temperature is observed as the average kink angle is less than 10 degree. This is credited to the insufficient Boltzmann excitation at low temperature regime. However, the geometric structure of the 50-CNA turns into chaos above the Peierls transition temperature where a sharper upturn is noticed above 500K in Figure 4. Using Boltzmann factor is not enough to explain the rapid increase of the average kink angle across the phase transition. The Peierls instability behaves as additional kinetic energy source which changes the period in the one-dimensional nanowire and presumably provides extra kinematic of atoms to form kinks [8,9,13]. As a result, the rigorous atomic motion owing to the Boltzmann factor in combination with the increase of entropy across the Peierls transition causes the carbon atoms to align more chaotically. The disorder in the 50-CNA starts to saturate above 800K as clearly seen in Figure 4 due to the stabilization of the thermal oscillations [11]. We confirm that the kink angle is unable to be influenced by Van Der Waal's force because it is too weak when compared to the strength of covalent bond in Figure 4. We proposed that the kink structure is non-negligible in the short carbon nanowire above the Peierls transition temperature. The kink structure should not be ignored in the DFT calculations especially at high temperature regime. For instance, the sharpness in Raman spectrum of short carbon chain should be decreased at high temperatures.

Can the kink structuring carbon nanowire absorb molecule effectively? The preliminary research on the carbon nanotube shows that the effective nuclear charge Z is increased due to curvature [3]. The kink structure in the short carbon chain creates numerous local curvatures and eventually the effective Z is expected to be enhanced. The average kink angle about 40 degree at in the 10 parallel 50-CNA at 1000K causes about 30% increase of the linear charge density from the positively charged lattice [14]. It may increase the efficiency to absorb negatively charged atoms at kinks and also open opportunity to absorb the disastrous gases emitted from combustion engines such as the negatively charged carbon monoxide [15,16].

**Conclusion:**

In summary, we have investigated the size effect on the kink structure in the monoatomic carbon chains via Monte Carlo method. Our results indicate that the crystalline periodicity should be present in the long carbon nanowire at any finite temperature but the alignment of the carbon atoms along the chain become more chaos above the Peierls transition temperature. Our simulation system opens the possibility to refine the atomic coordinates of the short carbon nanowire in the DFT simulation in order to achieve a more accurate theoretical prediction. It will be a valuable mission to keep investigating the thermal-induced kink structure in the carbon nanowire that may be used as the materials for chemisorption in future.

**Reference:**

[1] Srinivas Gadipelli, Zheng Xiao Guo, Graphene-based materials: Synthesis and gas sorption, storage and separation, Progress in Materials Science, Volume 69, Pages 1-60 (2015)

[2] Irena Yzeiri, Niladri Patra, and Petr Král, Porous carbon nanotubes: Molecular absorption, transport, and separation, The Journal of Chemical Physics 140, 104704 (2014)

[3] L. X. Benedict, V. H. Crespi, S. G. Louie, M. L. Cohen, Static conductivity and superconductivity of carbon nanotubes: Relations between tubes and sheets, Phys. Rev. B 52, 14935 (1995).

[4] Bruce H. Mahan, Rollie J. Myers, University Chemistry, Benjamin-Cummings Publishing Company, ISBN-13: 978-0201058338 (2000)

[5] Mingjie Liu, Vasilii I. Artyukhov, Hoonkyung Lee, Fangbo Xu, and Boris I. Yakobson, Carbyne from First Principles: Chain of C Atoms, a Nanorod or a Nanorope, ACS Nano, 7, (11), pp 10075–10082 (2013)

[6] Ivano E. Castelli, Paolo Salvestrini, and Nicola Manini, Mechanical properties of carbynes investigated by ab initio total-energy calculations, Phy Rev B, 85, 214110 (2012)

[7] C.H. Wong, J.Y. Dai, M.B. Guseva, V.N. Rychkov, E.A. Buntov, A.F. Zatsepin, Effect of symmetry on the electronic DOS, charge fluctuations and electron-phonon coupling in carbon chains, arXiv:1611.05584 (2016)

[8] Xiangjun Liu, Gang Zhang and Yong-Wei Zhang, Tunable Mechanical and Thermal Properties of One-Dimensional Carbyne Chain: Phase Transition and Microscopic Dynamics, J. Phys. Chem. C 119, 24156−24164 (2015)

[9] Alberto Milani, Matteo Tommasini, Daniele Fazzi, Chiara Castiglioni, Mirella Del Zoppo and Giuseppe Zerbi, First-principles calculation of the Peierls distortion in an infinite linear carbon chain: the contribution of Raman spectroscopy, Journal of Raman Spectroscopy, Volume 39, Issue 2, pp 164–168 (2008)

[10] L.Shi, P. Rohringer, K. Suenaga, Y. Niimi, J. Kotakoski, J.C. Meyer, H.M. Wanko, S. Cahangirov, A. Rubio, Z. J. Lapin, L. Novotny, P. Ayala & T. Pichler, Confined linear carbon chains as a route to bulk carbyne, Nature Materials, Vol 15, pp 634-640 (2016)

[11] J. R. Christman, Fundamentals of solid state physics, Wiley (1988)

[12] C.H.Wong, E.A. Buntov, V.N. Rychkov, M.B. Guseva, A.F. Zatsepin, Simulation of chemical bond distributions and phase transformation in carbon chains, Carbon, Volume 114, Pages 106-110 (2017)

[13] Jean-PaulPouget, The Peierls instability and charge density wave in one-dimensional electronic conductors, Comptes Rendus Physique Volume 17, Issues 3–4, Pages 332-356 (2016)

[14] John David Jackson, Classical Electrodynamics, Wiley, ISBN-13: 978-0471309321 (1998)

[15] Xinxing Shan,Yong Qian, Lifeng Zhu, Xingcai Lu, Effects of EGR rate and hydrogen/carbon monoxide ratio on combustion and emission characteristics of biogas/diesel dual fuel combustion engine, Fuel, Volume 181, 1 Pages 1050-1057 (2016)

[16] V.Arul Mozhi Selvan, R.B.,Anand, M.Udayakumar, Effect of Cerium Oxide Nanoparticles and Carbon Nanotubes as fuel-borne additives in Diesterol blends on the performance, combustion and emission characteristics of a variable compression ratio engine, Fuel, Volume 130, 15 Pages 160-167 (2014)